\documentclass[oneside,english]{amsart}
\usepackage[T1]{fontenc}
\usepackage[latin1]{inputenc}
\usepackage{geometry}
\usepackage{float}
\usepackage{graphicx}
\usepackage{amssymb}

\makeatletter
 \theoremstyle{plain}
\newtheorem{thm}{Theorem}[section]
  \theoremstyle{definition}
  \newtheorem{defn}[thm]{Definition}
  \theoremstyle{plain}
  \newtheorem{algorithm}[thm]{Algorithm}
  \theoremstyle{remark}
  \newtheorem*{acknowledgement*}{Acknowledgement}

\newcommand{\ra}{\rightarrow}

\usepackage{babel}
\makeatother
\begin{document}

\title{A Dual Algorithm for Non-abelian Yang-Mills coupled to Dynamical Fermions}
\author{J. Wade Cherrington$^1$}

\maketitle

\vspace*{-12pt}
\begin{center}

\small
$^1$Department of Applied Mathematics, University of Western 
Ontario, London, Ontario, Canada\\
e-mail: jcherrin@uwo.ca
\end{center}

\begin{abstract}
We extend the dual algorithm recently described for pure, non-abelian
Yang-Mills on the lattice to the case of lattice fermions coupled to Yang-Mills, by
constructing an ergodic Metropolis algorithm for dynamic fermions that is
local, exact, and built from gauge-invariant boson-fermion coupled 
configurations.
For concreteness, we present in detail the case of three dimensions, for the 
group $SU(2)$ and staggered fermions, however the algorithm readily generalizes
with regard to group and dimension.  The treatment of the fermion determinant makes
use of a polymer expansion; as with previous proposals making use
of the polymer expansion in higher than two dimensions, the critical question 
for practical applications is whether the presence of negative amplitudes 
can be managed in the continuum limit.

\end{abstract}

\section{Background}
Despite continued progress in algorithms and hardware, the inclusion of 
dynamical fermions in lattice gauge calculations continues to incur
significant computational expense. To motivate our proposal for a novel fermion 
algorithm, we briefly review how dynamical fermions are currently addressed.
Recall that dynamic fermions coupled to a gauge field on a $D$-dimensional hypercubic 
lattice for $D\geq2$ are governed by an action of the form
\begin{equation}
S[g_{e},\psi_{v},\overline{\psi_{v}}]=S_{G}[g_{e}]+S_{F}[g_{e},\psi_{v},\overline{\psi_{v}}],
\label{eq:fullaction}
\end{equation}
where the $g_{e}$ are valued in the gauge group $G$ at the edges of
the lattice and $\psi_{v}$ are the fermion fields defined at the
vertices of the lattice. 

Unlike gauge group variables, it is not practical to directly simulate
Grassmann variables on the computer. A common approach to dynamical fermion simulation starts 
by integrating out the fermion variables appearing in
$S_{F}[g_{e},\psi_{v},\overline{\psi_{v}}]$, 
to give a function of the gauge variables known as the fermion
determinant (its specific form is reviewed in Section 2).  The fermion 
determinant can be combined with the kinetic part of the gauge boson amplitude 
$e^{-S_{G}[g_{e}]}$ to give an effective action for the gauge variables
from which simulations on a computer can in principle proceed. However, the
fermion determinant  renders this effective action non-local~--- it couples 
together gauge variables that are arbitrarily distant in the lattice. This 
poses a considerable problem for the simulation, since computing
the change in the effective action due to a small change in any variable becomes 
very expensive, growing prohibitively with increasing lattice volume. A variety 
of algorithms have been devised to work with the fermion determinant;
a  description of some of the methods commonly employed can be found 
for example in \cite{DeDe}.

After reviewing the description of single component, staggered free fermions in terms
of self-avoiding polymers (as was done for example in~\cite{KST}), we review what
happens when a similar procedure is applied to multi-component fermion fields minimally 
coupled to gauge fields. In this case each polymer configuration corresponds to a 
Wilson loop functional; i.e. the trace of a product of representation matrices around 
the polymer. Because there is more than one component of the 
fermion fields in the non-abelian coupled case, the strict self-avoiding constraint of the
single component case is weakened; that is, for an $n$-component fermion, up to $n$ directed
polymer lines can enter and leave a given vertex. The picture has long been known --- it is essentially
that of a hopping parameter expansion of the fermion determinant, described for example in~\cite{Rothe}. 
Unlike many past applications, in the present case no cut-off in the power 
of the hopping parameter or otherwise is applied. Because we seek an exactly dual model, 
all polymers are included in the configurations considered. 

For each polymer diagram that arises in the free case, upon applying the duality transformation for
the group-valued field the result is a sum of configurations
consisting of all closed, branched, colored surfaces (spin foams) with open one-dimensional boundaries
defined by the polymer diagram.  The totality of spin foams associated with all polymer
diagrams (including the trivial empty polymer) defines the joint configuration space.
Crucially, local changes to the dual configurations (either polymer 
or surface structure) lead to local changes in the dual amplitude. 

The two theoretical inputs for this construction, 
a polymer decomposition of the fermion determinant (i.e. hopping parameter 
expansion as described in ~\cite{Rothe}) and a dual non-abelian model 
(e.g.~\cite{OecklPfeiffer} and references therein),
have been present in the literature for some time, and as we shall see the construction of 
the joint dual model at the formal level is a rather straightforward synthesis of these constituent models. 
However, unlike (the simplest implementations of) conventional lattice gauge simulations, 
finding \emph{any} practical algorithm for a dual model has proven somewhat non-trivial in
the non-abelian case for dimensions greater than two. The algorithm proposed here builds 
upon the dual non-abelian algorithm of \cite{CCK} that has recently been tested in the 
pure Yang-Mills sector. In addition to pure spin foam moves, we construct a set of moves 
that act on polymer structure and specify the type of vertex amplitudes that arise due 
to the charges carried by the polymer. Currently, an implementation of this algorithm
is being tested and will be reported on in a forthcoming work.

For context, it should be noted that a similar picture was present in the 
work of Aroca et al.~\cite{ArocaWorldsheet} and Fort~\cite{Fort}, which dealt with the abelian
case of $U(1)$ and proposed using a Hamiltonian that leads to 
a different Lagrangian formulation, where the ensemble is built from a restricted
subset of the configurations that arise in the Kogut-Susskind case.
In future work, we believe the non-abelian generalization of~\cite{ArocaWorldsheet} may be a 
very interesting alternative to the Kogut-Susskind formulation used here,
particularly if the imbalance between negative and positive amplitudes and 
the reduction of species doubling described in~\cite{ArocaWorldsheet} 
can be carried over to the non-abelian case.  

The outline of the paper is as follows. In Section~2, we review the origin
of the fermion determinant and discuss its expansion in terms of 
polymers.
In Section~3, we briefly review the dual computational framework for non-abelian,
pure Yang-Mills theory on the lattice. In Section~4, we show a natural
way to to combine these frameworks and formulate ergodic moves for the coupled
fermion-boson system. In Section~5, we offer some conclusions and
describe our program for ongoing numerical work based on this algorithm and
its extensions. Appendix~A describes the vertex amplitudes that arise in the
joint case, while Appendix~B expands on Section~2 to describe the trace
structure of polymers with multiply occupied vertices.


\section{Polymer description of fermions on the lattice}
In this section, we start from a conventional lattice discretization of free 
fermions, following \cite{M&M} in essentials and notation. In the usual manner,
the fermion determinant is arrived at by exact integration of the Grassmann 
variables.  Following \cite{KST}, the fermion determinant is then expanded into
states, each of which is represented by a family of \emph{disjoint, closed
oriented loops}, including trivial and degenerate ``loops'', the \emph{monomers} and \emph{dimers},
respectively. 
A typical polymer configuration (in the $D=2$ massive case) is illustrated~\footnote{Figure 1 shows
the fermion state on part of a larger lattice to which periodic boundary conditions are applied.} 
We now review explicitly how the fermion determinant and polymer picture come about. 
Note this section is purely for pedagogical purposes and to fix notation to be used later;
those familiar with the hopping parameter expansion of the fermion determinant can safely
skip it. 

\begin{figure}
\includegraphics[scale=0.45]{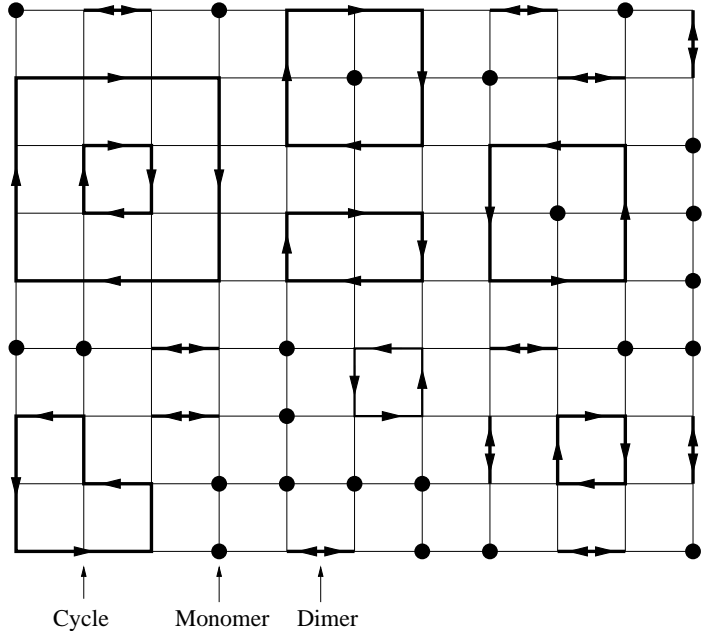}
\caption{Part of a typical configuration in the polymer expansion of the fermion 
determinant in two dimensions (massive case). }
\label{fig:2dexample}
\end{figure}

\subsection{Kogut-Susskind staggered fermions --- free case }
To illustrate the concept and introduce terminology, we treat a single species 
of fermions with no additional indices.We start with the naieve lattice field 
action for free staggered fermions
\begin{equation}\label{lfermions}
S = \sum_{x\in V}\overline{\Psi}_{x}\left(\gamma_{\mu}\partial_{\mu}+m\right)\Psi_{x},
\end{equation}
where $\gamma_{\mu}$ are the (Euclidean) Dirac matrices and $V$ is the set
of lattice vertices. 

Using the central difference for the partial derivative, this becomes 
\begin{equation}
S=\sum_{x\in V}a^{D}\left\{ m(\overline{\Psi}_{x}\Psi_{x})-\frac{1}{2a}\sum_{\mu=1}^{D}
(\overline{\Psi}_{x}\gamma_{\mu}\Psi_{x+\hat{\mu}}-\overline{\Psi}_{x+\hat{\mu}}\gamma_{\mu}\Psi_{x})\right\},
\label{eq:freefermionundiag}
\end{equation}
where $D$ is the dimension\footnote{In three dimensions, the continuum
limit of the staggered fermion action contains flavours correponding
to two inequivalent representations of the Dirac algebra~\cite{BB}.}
of the hyper-cubic lattice, $m$ is the mass, $a$ is the lattice spacing,
and $\mu$ labels one of the $D$ directions of the lattice; $\hat{\mu}$ is the
unit lattice vector associated to the $\mu$th direction.
Following \cite{M&M}, we change to a basis which diagonalizes the gamma
matrices, rewriting the free action as
\begin{equation}
S=\sum_{x\in V}\left\{
M(\overline{\psi}_{x}\psi_{x})-K\sum_{\mu=1}^{D}\alpha_{x\mu}\left(\overline{\psi}_
{x}\psi_{x+\hat{\mu}}-\overline{\psi}_{x+\hat{\mu}}\psi_{x}\right)\right\},
\label{eq:freefermiondiag2}
\end{equation}
with $\alpha_{x\mu}\equiv(-1)^{x_{1}+\cdots+x_{\mu-1}}$ for $\mu\in\{1,2,...,D\}$
where the $x_{i}$ are the components of the lattice site four-vector
and $\frac{M}{2K}=ma$. The edge dependent sign factor $\alpha_{x\mu}$ arises from 
the chosen diagonalization.

To express the result of integration over the Grassmann variables it
is convenient to introduce the \emph{quark matrix} $Q$, defined in terms
of lattice regularized action as 
\begin{equation}
S_{F}[g_{e},\psi_v,\overline{\psi}_v]=\sum_{x,y\in V}\overline{\psi}_{y}Q[g_{e}]_{yx}\psi_{x}.
\label{eq:qmatrix}
\end{equation}
For later reference, we include dependence on the gauge degrees of
freedom in our definition of $Q$. We next apply the well known result \cite{M&M} that the
Grassmann integral over the fermion fields at every vertex evaluates to
\begin{equation}
\int\left(\prod_{v\in V}d\psi_{v}d\overline{\psi}_{v}\right)e^{-S_{F}[g_{e},\psi_v,
\overline{\psi}_v]}=\int\left(\prod_{v\in V}d\psi_{v}d\overline{\psi}_{v}\right)
e^{{-\sum}_{x,y\in V}\overline{\psi}_{y}Q[g_{e}]_{yx}\psi_{x}}=\det Q[g_{e}],
\label{eq:detidentity}
\end{equation}
the determinant of the quark matrix.

Next we recall the continuum form of the action for massive fermions coupled
to a gauge field. In the massive case, the quark matrix can be written as
\begin{equation}
Q_{yx}=M\delta_{yx}-K_{yx},
\label{eq:QuarkMatrix}
\end{equation}
where $M$ is the fermion mass and $K_{xy}$ is the \emph{hopping
matrix} that is non-zero for nearest neighbor pairs $(x,y)$. By
inspection of (\ref{eq:freefermiondiag2}), we write the quark 
matrix as
\begin{equation}\label{Qdetail}
Q_{yx}= M\delta_{yx}-K\sum_{\mu=1}^{D}\alpha_{x\mu} 
( \delta_{y,x-\hat{\mu}}-\delta_{y-\hat{\mu},x}).
\end{equation}
We now apply a well known identity for the determinant of (any) matrix $Q$,
\begin{equation}\label{eq:perms}
\det Q=\sum_{\pi}\text{sgn}(\pi)\prod_xQ_{x\pi(x)},
\end{equation}
where $\pi$ ranges over the set $\Pi$ of all permutations of the indices of $Q$ and $\text{sgn}(\pi)$ 
is the sign of the permutation. In the case of the quark matrix $Q$, the matrix indices
being permuted correspond to vertices of the lattice.  Thus, one is led to consider
the product $\prod_xQ_{x\pi(x)}$ for every permutation of lattice vertices.  

The next step is to recognize that every permutation $\pi$ can be decomposed into
a composition of  disjoint, non-trivial \emph{cyclic} permutations $\pi^c$, $\pi = \prod_i\pi_i^c$. 
For a matrix with all entries non-zero, these permutations may involve sets of vertices that are
arbitrarily separated on the lattice. However, the quark matrices that arise 
in practice have a very specific structure (originating in the lattice discretization 
from nearest neighbor approximations of the derivative operator); the only non-zero
matrix elements consist of nearest-neighbor pairs (and in the massive case, on-diagonal).
Thus, the non-zero contributions can be analyzed as follows.

A given permutation $\pi$ affects any vertex trivially (the vertex is sent to itself) 
or as part of a non-trivial cyclic permutation. In the massive case, vertices that are
permuted trivially give \emph{monomer} factors equal to the mass $M$. 
In the massless case where diagonal entries vanish, any trivial permutation will lead
to a vanishing contribution; thus every vertex must participate in a cyclic 
permutation in the massless case.  

For the non-trivial permutations, it is useful to distinguish two cases that a
vertex may participate in.  Permutations that swap a pair of neighboring vertices 
are referred to as \emph{dimers}; all other non-trivial permutations consist of
non-trivial loops of edges on the lattice; we shall refer to these as \emph{cycles}. 
Given these observations on the structure of $Q$, we can now write an expression
for $\det Q$ in more explicit detail as 
\begin{equation}
\det Q_{yx} =  \sum _{\pi \in \Pi} \text{sgn}(\pi)M^{N_m}\prod_{(xy)\in p}K_{xy}.
\label{eq:polymers}
\end{equation}
Where $N_m$ are the number of monomers. The product is over all directed edges $(xy)$ that
are part of a dimer or cycle of the permutation.

\subsection{Coupled case} \label{sse:coupled}

To couple fermions to the gauge fields, the ordinary derivative is replaced by the covariant
one, thereby introducing the gauge variables $g_{e}$ which act on the fermions through the
matrices of the representation corresponding to the charge of the fermion. For specificity, 
we will consider fermions charged in the representation of $G$ labelled by $c$. One can
 show~\cite{M&M} that the lattice action for staggered fermions coupled to the gauge field becomes
\begin{equation}\label{eq:coupled}
S = \sum_x \left( M(\overline{\psi}_x \psi_x)+K\sum_{\mu}^{D}\alpha_{x\mu}
\left( \overline{\psi}_xU_{x\mu}^\dagger\psi_{x+\hat{\mu}} - \overline{\psi}_{x+\hat{\mu}}
U_{x\mu}\psi_x\right)\right ).
\end{equation}
In contrast to the (simplified one component) free case, there is in general 
a vector possessing  multiple Grassmann variable components at each vertex, and matrices $U(g_e)$ 
that act non-trivially on this vector. In terms of our permutation expansion, 
the quark matrix now takes on component as well as vertex labels as follows:
\begin{equation}
S_{F}[g_{e},\psi_v,\overline{\psi}_v]=\sum_{x,y\in V}\overline{\psi}_{y}^jQ[g_{e}]_{yx}^{ji}\psi_{x}^i.
\label{eq:quarkmatrixmulti}
\end{equation}
Comparing~(\ref{eq:quarkmatrixmulti}) to~(\ref{eq:coupled}) we identify the multicomponent
quark matrix as 
\begin{equation}\label{eq:Qdetailmulti}
Q_{yx}^{ji}[g_e]= M\delta_{yx}-K\sum_{\mu=1}^{D}\alpha_{x\mu} \left( (U^{\dagger}_{x\mu})^{ij}
\delta_{y,x-\hat{\mu}}-(U_{x\mu})^{ij}\delta_{y-\hat{\mu},x}\right).
\end{equation}
The $g_e$ dependence is through the representation matrices $U_e$, where
$e$ is labelled in one of the two conventions introduced above. 
The determinant formula can again be applied; upon doing so, the expansion 
into permutations takes on the form
\begin{equation}\label{eq:perms_vec}
\det Q[g_e]=\sum_{\pi}\text{sgn}(\pi)\prod_{x,i}Q_{x\pi_{V}(x,i)}^{i\pi_{I}(x,i)}.
\end{equation}
Observe that the permutations $\pi$ now act on both vertex and component indices of 
$Q$; the action of $\pi$ on indices and vertices can be separated into the maps
$\pi_I$ and $\pi_V$, respectively.  As in the free case, the locality structure allows one to 
identify non-vanishing, polymer-like contributions.
For $\pi$ where vertex index mapping is trivial, the only non-vanishing entries
are those for which $\pi_I(x,i)=i$ as the massive term is an inner product with
no cross-terms. Thus, the monomers contribute factors $Mn$, where
$n$ are the number of components of the fermion vectors.

As in the single-component case, for vertices that participate in non-trivial permutations,
 one still has that only permutations which move $x \rightarrow \pi(x)$ where $\pi(x)$ is a nearest neighbor
are non-vanishing. However, due to the presence of multiple components to the fermion field, 
permutations can shift both components at a vertex simultaneously. 

Configurations which involve non-trivial shifts of more than a single component 
(multiply occupied vertex contributions) are discussed in the Appendix B.  In the 
remainder of this section, we will restrict ourselves to the case where only a single 
component participates in the shift, to illustrate in a simple setting how the trace of a product of $U_{e}$ matrices comes about.

\begin{figure}[tbp]
\includegraphics[scale=0.45]{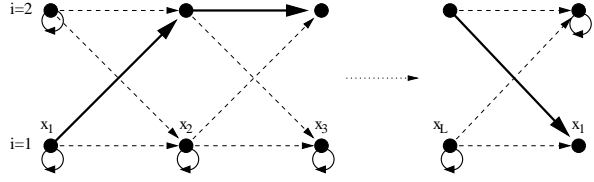}
\caption{ Graphical representation of one (solid) of all possible
permutations (dashed) associated with a cycle of vertices of length $L$. }
\label{fig:paths}
\end{figure}

By inspection of~(\ref{eq:coupled}), we see for a given nearest neighbor vertex shift, 
\emph{all possible permutations of indices are allowed}, since in the general case $U^{ij}_{x\mu}$ has all non-vanishing entries. 
To continue our analysis we factor the permutation into a part that acts on vertices
$\pi_V$ (these correspond to the dimers and cycles of the free case) 
and a part that acts on component indices $\pi_I$. We now write the fermion determinant
as
\begin{equation}\label{eq:perms_vec2}
\det Q[g_e]=\sum_{\pi_V}\text{sgn}(\pi_V)\sum_{\pi_I}\prod_xQ_{x\pi_V(x)}^{i\pi_I(x,i)}+\text{\emph{(multiply occupied vertex contributions)}}.
\end{equation}
Focusing our attention on the first term, we note that only one component per vertex
is shifted in each of the products of the sum (multiple component shifts at a vertex are precisely
what is included in the second term, discussed in Appendix B).  For a given $\pi_V$, we 
can represent $\pi_I$ as an ordered sequence of arrows through the discrete $n$-point space 
living above every vertex acted on by $\pi_V$; see Figure~\ref{fig:paths}.
Note that for a given dimer or cycle $\pi_V$, all possible index sequences correspond
to permutations of the same order (thus $\text{sgn}(\pi_V)$ can be factored out). 
The final step in the analysis is to recognize that the sum over all paths
is simply the trace of the matrix product of representation matrices around the cycle.

We define $D_1$ as the restriction of $\det Q[g_e]$ to permutations involving singly occupied vertices; 
that is, the first term of~(\ref{eq:perms_vec2}). $D_1$ can be constructed out of loops 
with a single associated trace as follows:
\begin{eqnarray}\label{eq:perms_detail}
D_{1} &=&\sum_{\pi_V}\text{sgn}(\pi_V)\sum_{\pi_I}\prod_xQ_{x\pi_V(x)}^{i\pi_I(i)}
\\  &=& \nonumber \sum_{\pi_V}\text{sgn}(\pi_V)(nM)^{N_m}K^{N_e}\left(\prod_{(xy)\in \pi_V}
\alpha_{(xy)} \right) U_{(x_1x_2)}^{i_1i_2}U_{(x_2x_3)}^{i_2i_3}U_{(x_3x_4)}^{i_3i_4}
\cdots U_{(x_Lx_1)}^{i_Li_1} 
\\  &=& \nonumber \sum_{\pi_V}\text{sgn}(\pi_V)(nM)^{N_m}K^{N_e}\left(\prod_{(xy)\in \pi_V}
\alpha_{(xy)} \right) \text{Tr}\left(\prod_{(xy) \in \pi_V} U_{(xy)}\right),
\end{eqnarray}
where $N_e$ is the total number of edges where vertices are shifted in the permutation.
In the second line, the Einstein summation convention for repeated indices is
used. Observe that $(xy)$ denotes an oriented edge. Depending on whether $(xy)$
is along or opposing the canonical orientation, one has either a product 
of $U_{(xy)}$ or $U_{(yx)} = U^{\dagger}_{(xy)} $. 
The visualization of a typical polymer configuration on a $D=3$ lattice (including
those with multiply occupied vertices) appears in Figure~\ref{fig:fermionvisual} below.

So far our discussion has been generic with regard to group, dimension, and fermion
charge; the only major choice has been staggered fermions rather than an alternative lattice
discretization. In the remainder of this work, we restrict our attention to $D=3$, $G=SU(2)$,
and $n$-component massive fermion fields charged with half-integer spin $c$ and minimally coupled to $SU(2)$.
\begin{figure}
\includegraphics[scale=0.55]{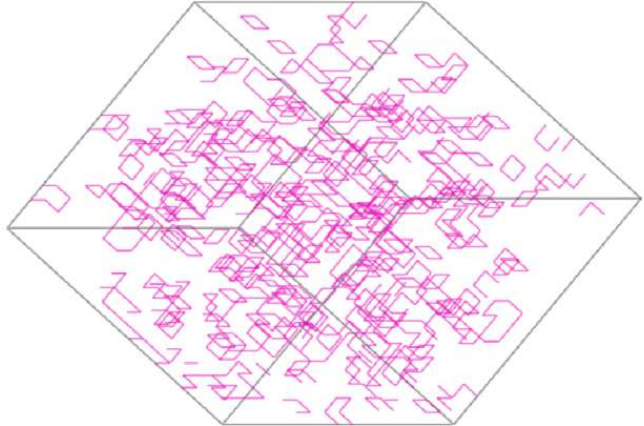}
\caption{Visualization of 2-component fermion field on $16^3$ lattice. Polymers that appear
to have open ends close on the opposite side of lattice due to periodic boundary conditions.}
\label{fig:fermionvisual}
\end{figure}


\section{Spin Foam Description of Pure Gauge Theory on the Lattice}
Having described what we shall refer to as the free (no coupling to gauge fields) fermion partition function 
 in terms of closed lattice polymers, in this section we briefly review the dual 
formulation of pure (no coupling to fermions) Yang-Mills, which leads to closed, 
colored, branched lattice surfaces.
We shall see in the next section that these pictures naturally combine to give the 
full interacting partition function in terms of a space of coupled configurations.

It can be shown (see for example~\cite{Conrady, OecklPfeiffer}, and~\cite{CCK} for detail 
on $G=SU(2)$ in three dimensions) that starting from the lattice discretized action 
for pure Yang-Mills
\begin{equation}\label{eq:conventional}
	\mathcal{Z}_B = \int \prod_{e\in E} dg_e\, e^{-\sum_{p\in P} S(g_e)},
\end{equation}
one can transform to a spin foam formulation expressing the partition function
in terms of dual variables as follows:
\begin{equation}
\mathcal{Z}_B =\sum_{j}\left(\sum_{i}
	\prod_{v\in V}{18j}^v(i_{v},j_{v})
	\prod_{e\in E} N^e(i_e,j_e)^{-1}\right)
\left(\prod_{p\in P}e^{-\frac{2}{\beta}j_{p}(j_{p}+1)}(2j_{p}+1)\right).
\label{eq:statesum}
\end{equation}
Here $V$, $E$, and $P$ denotes the vertices, edges, and plaquettes of
the lattice, respectively. The summations over $i$ and $j$ range over 
all possible edge and plaquette labellings, respectively. A 
plaquette labelling $j$ assigns an irreducible
representation of $SU(2)$ to each element of $P$. These representations
are labelled by non-negative half-integers (we will denote this set 
by $\frac{1}{2}\mathbb{N}$), also referred to as \emph{spins};
a labelling $j$ is thus a map $j\colon P\to\frac{1}{2}\mathbb{N}$.
In the $SU(2)$, $D=3$ case, edges are also labeled by half-integer
representations. thus an edge labelling is a map $i\colon E\to\frac{1}{2}\mathbb{N}$.

Following~\cite{CCK}, we define a (vacuum) \emph{spin foam} configuration
as one summand in~(\ref{eq:statesum}), i.e. a labelling of both 
plaquettes and edges by spins and intertwiners, respectively. The amplitude assigned to a spin foam
factors into a local product of amplitudes.  The vertex amplitude ($18j$ symbol) depends on 
the 12 plaquettes and 6 edges incident to a vertex; the $N_e$ factors
depend on the 4 plaquettes and the intertwiner labelling of an edge, and there
is a product of local plaquette factors.

The locality of the spin foam formulation was applied in~\cite{CCK} to
perform computations that were verified against conventional methods.

\begin{figure}
\includegraphics[scale=0.55]{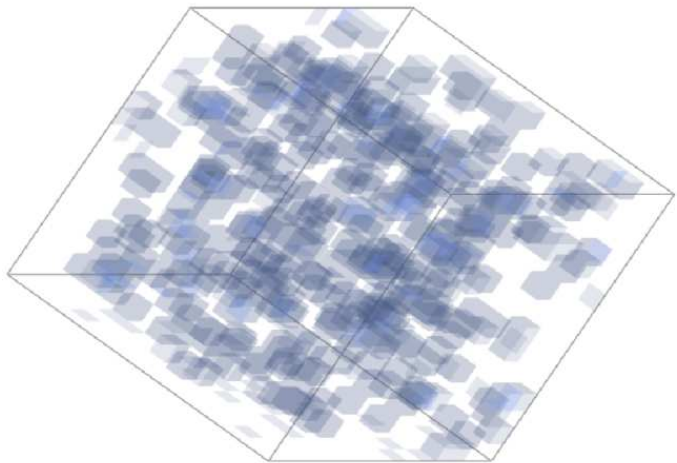}
\caption{Visualization of pure Yang-Mills vacuum on $16^3$ lattice.}
\label{fig:foamvis}
\end{figure}


\section{Dual Fermion-Boson Simulations}
In this section we describe how the dual pictures of lattice fermions and gauge bosons 
presented in the previous two sections can be combined to form a joint dual partition function,
built up of gauge-invariant configurations with discrete occupancy and representation
labels. 

\subsection{The Joint Partition Function}
Using the action $S[g_{e},\psi_v,\overline{\psi_v}]$ for the full theory we write the 
partition function as follows:
\begin{eqnarray}
\mathcal{Z}_J&=&\int\left(\prod_{e\in E}dg_{e}\right)\int\left(\prod_{v\in V}d\psi_{v}
d\overline{\psi}_{v}\right)e^{-S_{F}[g_{e},\psi_{v},\overline{\psi_{v}}]}e^{-S_{G}[g_{e}]} \\ \nonumber
&=&\int\left(\prod_{e\in E}dg_{e}\right)\det Q[g_{e}]e^{-S_{G}[g_{e}]},
\end{eqnarray}
where we have integrated out the fermionic variables to get the fermion
determinant. We next use the polymer expansion for the determinant in
the case of gauge-coupled $Q$, as described in Section~\ref{sse:coupled}:
\begin{eqnarray}
\mathcal{Z}_J&=&\int\left(\prod_{e\in E}dg_{e}\right)\det Q[g_{e}]e^{-S_{G}[g_{e}]} \\ \nonumber
&=& \int\left(\prod_{e\in E}dg_{e}\right)
\sum_{\gamma}\text{sgn}(\gamma)(nM)^{N_m}K^{N_K}\left(\prod_{(xy)\in \gamma}
\alpha_{(xy)} \right) \text{Tr}\left(\prod_{(xy) \in \gamma} U_{(xy)}\right)
e^{-S_{G}[g_{e}]}. \\ \nonumber
\end{eqnarray}
Here $N_K$ is the number of $K$ factors (one per unit of edge occupancy) in the polymer configuration;
the definition of polymer configuration for $D=3$, $G=SU(2)$ is given in Section 4.3.1.
In the second line, we have substituted the polymer expansion for fermion determinant.
Next, we recall the form of the character expansion (see~\cite{CCK} and references therein) 
for the amplitude based on the heat kernel action at a plaquette $p$,
\begin{equation}
	e^{-S_p(g)} = \frac{1}{K(I,{\textstyle \frac{\gamma^2}{2}})}
		\sum_j (2j+1) e^{-\frac{\gamma^2}{2}j(j+1)} \chi_j(g),
		\quad {\textstyle j=0,\frac{1}{2},1,\ldots}
\end{equation}
Substituting the character expansion into the previous equation, we have
\begin{eqnarray}
\mathcal{Z}_J &=& \int\left(\prod_{e\in E}dg_{e}\right)
\sum_{\gamma}\text{sgn}(\gamma)(nM)^{N_m}K^{N_K}\left(\prod_{(xy)\in \gamma}
\alpha_{(xy)} \right) \text{Tr}\left(\prod_{(xy) \in \gamma} U_{(xy)}\right)
 \\ \nonumber
&\times& \prod_{p \in P}\sum_{j_p} (2j_p+1) e^{-\frac{\gamma^2}{2}j_p(j_p+1)} \chi_j(g),
\end{eqnarray}
where an overall constant factor of $K(I,\frac{\gamma^2}{2})$ 
per plaquette has been discarded. We show in Appendix A that the group integrals 
over products of traces and characters in each term of the character expansion
can be evaluated exactly in terms of charged 18j symbols, provided a 
sum over intertwiner labels is made at each edge. Using the vertex and edge amplitudes
of Appendix A, we can exhibit the joint dual partition function as
\begin{eqnarray} \label{eq:jointpartfunc}
\mathcal{Z}_J &=& \sum _{\gamma \in \mathcal{P}}\sum_{j}\sum_{i}
	\left(s(\gamma)\prod_{v\in V}\overline{18j^v}(i_{v},j_{v},\gamma)
	\prod_{e\in E} \overline{N^e}(i_e,j_e,\gamma)^{-1}\prod_{p\in P}e^{-\frac{2}{\beta}j_{p}(j_{p}+1)}(2j_{p}+1)\right),
\end{eqnarray}
where $s(\gamma)\equiv\text{sgn}(\gamma)\prod_{(xy)\in \gamma}\alpha_{(xy)}(nM)^{N_m}K^{N_K}$ 
combines the two sign factors and a product of $M$ and $K$ factors.
As we shall see in the next section, joint configurations associated to a 
polymer $\gamma$ carry in general three rather than a single intertwiner label $i_e$ for 
each edge belonging to the polymer; $i$ here ranges over all the intertwiner labels.

Although the overall dual amplitude is still a product of local amplitudes as in 
the pure Yang-Mills case, the presence of the Wilson loop functionals associated 
to non-trivial polymers requires the vacuum vertex and edge amplitudes to be 
modified in a way that we define in Appendix~A; the result is a product
of modified $\overline{18j^v}$ symbols and edge amplitudes $\overline{N^e}$ that 
are charged according to the polymer content $\gamma$ of the configuration.

The joint ensemble that results here can be viewed as a generalization
of the usual definition of spin foams to include one-dimensional
structure corresponding to the presence of fermionic charge. 
For a given polymer, there is a sum over all spin foams satisfying
admissibility, which is modified at the polymer edges. 
From the worldsheet point of view~\cite{ConradyKhavkine}, a polymer loop
acts as the source or sink of $2c$ fundamental sheets, where $c$ is the
half-integer charge of the fermion.


\subsection{The Joint Fermion-Boson Configurations}
In this section we define explicitly the set of configurations that include
all those that give non-zero contributions\footnote{Due to exceptional zeros 
there may be configurations that are admissible by
the conditions defined in this section, but are nonetheless zero. As in the
pure Yang-Mills case~\cite{CCK}, we assume exceptional zeros are sufficiently
isolated that ergodicity on admissibles is equivalent to ergodicity on non-zero
configurations. } to the joint dual partition function.

Specifically, for a given polymer $\gamma$, we introduce definitions that will allow us to 
characterize the set of spin foam configurations (plaquette colorings)
that are admissible in the presence of $\gamma$.

As in the pure Yang-Mills case~\cite{CCK}, we assume a splitting has been made for 
each edge with $j_1$ and $j_2$ on one side and $j_3$ and $j_4$ on the
other. Because of the presence of charges $c_1$ and $c_2$, the intertwiner is generally
6-valent and three splittings have to be made. For discussing edge admissibility, we 
assume the splitting is such that $c_1$  and $c_2$ are in the middle of the channel
as shown in Figure~\ref{fig:splitv}. Less symmetric splittings are possible, but we restrict
our attention to this case in the following.  

Let $\gamma$ denote a polymer configuration of charge $c$.
We define the set of \emph{$\gamma$-admissible plaquette configurations} as
those configurations whose labellings satisfy \emph{$c$-edge admissibility}
at every edge in the lattice, where $c$-admissibility is defined as follows:

\begin{defn}[$c$-edge admissibility]
The spins assigned to plaquettes incident to an edge are said to 
be \emph{$c$-edge admissible} if the \emph{parity}
and \emph{triangle inequality} conditions are satisfied. The charge insertions $c_1$ and $c_2$ may
be 0 or $c$, according to whether the edge is uncharged, singly charged, or doubly charged.  
Writing $j_1$, $j_2$, $j_3$ and $j_4$ for the four spins incident to a given edge, these conditions are
\begin{enumerate}
\item \textbf{Parity}: \[
j_{1}+j_{2}+j_{3}+j_{4} + c_{1} + c_{2} \text{ is an integer.}\]

\item \textbf{Triangle Inequality}: for each permutation $x\equiv(x_1,x_2,x_3,
x_4,x_5,x_6)$ of the charge and spin variables $\left(c_1,c_2,j_1,j_2,
j_3,j_4\right)$ we have
\[
 x_1 + x_2 + x_3 + x_4 +x_5\geq x_6 .
\]
\end{enumerate}
These conditions are equivalent to the existence of a non-zero
invariant vector in the $SU(2)$ representation 
$c_1\otimes c_2\otimes j_1\otimes j_2 \otimes j_3\otimes j_4$.
\end{defn}

The allowed range of intertwiner labels $i_1$, $i_c$ and $i_2$ depend on
the incident spin labels, on each other and on $c$ through the vertices 
where the charges $c_1$ and $c_2$ enter the diagram.  
We now state the condition for and admissible spin foam (plaqeutte
and edge intertwiner labelling) in the presence of an arbitrary
polymer $\gamma$:

\begin{figure}
\includegraphics[scale=1.0]{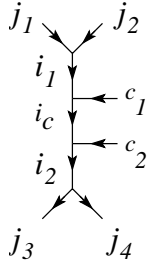}
\caption{Symmetric splitting of a 6-valent $SU(2)$ spin network vertex.}
\label{fig:splitv}
\end{figure}

\begin{defn}[$\gamma$-admissible spin foam]\label{defn:admissible-spin-foam}
A spin foam is \emph{$\gamma$-admissible}  if and only if for every edge $e\in E$:

\begin{enumerate}
\item  The plaquettes incident to $e$ are 
$c$-edge admissible in the sense of Definition 4.1, with $c_1$ and
$c_2$ assigned depending on the occupancy of $e$ by $\gamma$.
\item Each vertex of Figure~\ref{fig:splitv} is admissible. Explicitly, the following conditions are simultaneously satisfied:
\[
i_{2}+j_{1}+j_{2},i_1+i_c+c_1,i_2+i_c+c_2 \text{ and } i_{1}+j_{3}+j_{4} \text{ are integers} 
\]
and
\[
i_1 \in [|j_1-j_2|,j_1+j_2]\cap[|i_c-c_1|, |i_c+c_1|],
\]
\[
i_c \in [|i_1-c_1|,i_1+c_1]\cap[|i_2-c_2|, |i_2+c_2|],
\]
\[
i_2 \in [|j_3-j_4|,j_3+j_4]\cap[|i_c-c_2|, |i_c+c_2|].
\]
\end{enumerate}

For a given polymer $\gamma$, we denote the set of $\gamma$-admissible spin foams by $\mathcal{F}^{A}_{\gamma}$.

\end{defn}

\subsection{The Joint Moves and Algorithm} In this section we define moves that transform
from one joint configuration to another.  Together, they are ergodic and obey detailed balance and can
thus be used in a Metropolis or other Markov chain Monte Carlo algorithms. 

\begin{defn}[Pure spin foam move]\label{defn:sfmove}
A pure spin foam move consists of a single application of the \emph{cube}, \emph{edge}, or
\emph{homology} move. In terms of their effect on plaquette spins, these moves are as 
defined in~\cite{CCK}.  However, their effect on intertwiner labels needs to be generalized 
to account for the extra intertwiner labellings introduced by polymers.
\end{defn}

Because polymer moves require simultaneous changes in spin foam structure, we use the term
``pure'' to distinguish spin moves that leave the polymer structure unchanged.
We now describe the generalization of each pure spin foam move to account for
extra intertwiner labels. 

\begin{defn}[Generalized cube move] As described in Definition 2.4 and Appendix
A.3 of~\cite{CCK}.
With reference to compatible intertwiner moves of Type A, no changes in intertwiner
labels are necessary. For Type B edges, all three labels are increased or decreased
by the same half-unit of spin.
\end{defn}

\begin{defn}[Generalized edge move] As described in Definition 2.5 of~\cite{CCK}.
Rather than changing the single intertwiner label, the three intertwiner labels 
$i_1(e)$, $i_2(e)$ and $i_c(e)$ are each randomly changed by $-2$,$0$, or $2$ (to preserve 
parity) units of spin. 
\end{defn}

\begin{defn}[Generalized homology move] As described in Definition 2.6 of~\cite{CCK}, but 
all three intertwiner labels are increased or decreased by one half-unit of spin.
\end{defn}

\subsubsection{Polymer moves}
In Section~\ref{sse:coupled}, we saw how sums over permutations in $\Pi$ can be encoded
into traces of matrix products, ordered according to the orientation of the permutation.
Our example was restricted to singly occupied vertices, and is generalized in Appendix B.
Combining these cases, we see that the sum over all permutations in $\Pi$ can be represented
by traces of products of matrices, if we include diagrams corresponding to all possible routings
at multiply occupied vertices.  This new set of objects, oriented diagrams with routings at multiply
occupied vertices, we refer to as \emph{polymers}, and denote by $\mathcal{P}$. It is important
to distinguish the polymers from their finer-grained constituents, the permutations $\Pi$, as
polymers can be coupled naturally into the spin foam partition function, whereas individual
permutations cannot be using the methods here. 

In this section we present a set of moves that are ergodic on the
space of polymers $\mathcal{P}$ on a 3-dimensional hypercubic lattice.
The polymer states at an edge in the $2$-component case considered here are as follows. Assuming
a global orientation has been selected for the edges, an edge can be unoccupied, singly occupied,
or doubly occupied with the occupied cases carrying both positive and negative orientation.
The occupancy data at each edge can thus be assigned from the set $\{-2,-1,0,1,2\}$.
We shall use the term ``line of flux'' interchangeably with directed polymer line.

\begin{defn}[Plaquette move]
A plaquette and plaquette orientation (clockwise or counterclockwise) is randomly selected. 
To each edge, a delta occupancy of $+1$ or $-1$  (with signs given according to plaquette orientation)
is assigned, and added to the present occupancy.  If the resulting occupancy on any edge has 
magnitude greater than 2, the move is immediately rejected.  At each multi-valent vertex, a
choice of routing is made with equal probability. A proposed move that removes occupancy of edges
incident on multiply occupied vertices must make further random choice of routing that matches the routing present or
be rejected. This is necessary to preserve detailed balance.
If the move is not rejected, the spin of the selected plaquette is randomly decreased or increased by $c$ to 
satisfy parity. Because a change in the polymer occupancy forces a change in both the plaquette spin and the charge 
structure at an edge, the affected intertwiner labels (at each edge of the affected plaquette) must 
change in a way that is compatible; if not the result will be immediately $c$-edge inadmissible
for the new charging. 
\end{defn} 

\begin{figure}
\includegraphics[scale=0.52]{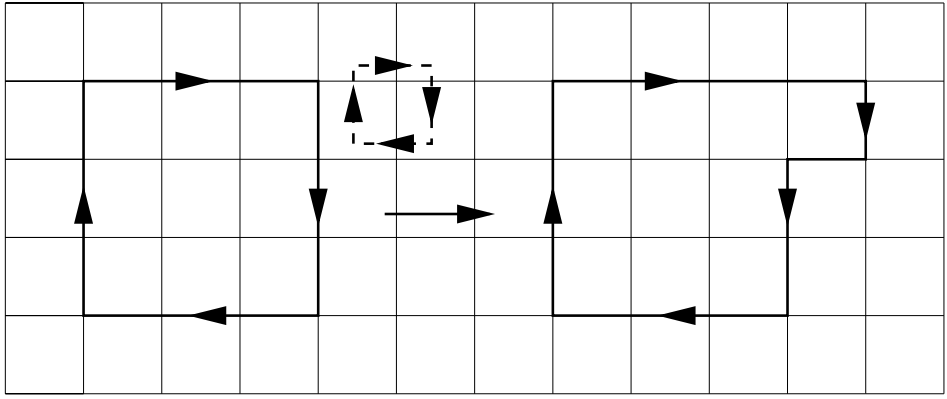}
\caption{Stretching of a loop by a plaquette move}
\label{fig:pmoves}
\end{figure}

This is the most fundamental polymer-changing move, and connects a very large region of the space of polymers contributing
to the fermion determinant. One can see trivially that the plaquette move can create fundamental
loops of either orientation when applied to ``empty'' space (plaquettes of zero occupancy edges).
Figure~\ref{fig:pmoves} illustrates how the plaquette move can deform an existing cycle.
The same plaquette move of opposite orientation would lead to one of
multiple routings with a doubly occupied edge, as shown by Figure~\ref{fig:pchoices} in Appendix~B.

Because an equally weighted routing choice is made amongst several alternatives when the plaquette
move of (for example) Figure~\ref{fig:pchoices} is made, the move that reverses a particular routing to get
the initial loop back should occur with proportional probability, in order to satisfy detailed balance.
This will unfortunately lead to a lowered acceptance rate, particularly in regimes with high occupancy. 

To identify intertwiner moves compatible with a plaquette move, one must give 
the choice of splitting (grouping of $j_1, \ldots ,j_4$ into pairs) around the edges of a plaquette, in 
the same way that intertwiner moves compatible with a  cube move depends on the splitting (see A.3 of~\cite{CCK}).  
For conciseness and generality, we have made the present definition of the algorithm splitting independent. 
Forthcoming work on numerical simulations with this algorithm will evaluate alternative splittings and describe 
the appropriate compatible moves.

\begin{defn}[Global circle move] For integer valued charge $c$, a global circle
moves adds a single line of charge and a minimal cylinder of $\frac{c}{2}$ charged plaquette spins spanning the lattice.
The origin and orientation of the cylinder is randomly selected to lie on a plaquette of one
of three orthogonal lattice planes through the origin.

For half-integer $c$, two lines of charge spanning the lattice are added and
a minimal surface consisting of a line of $c$-charged plaquettes is added between the
charge lines.  The position and orientation of the sheet is randomly selected to lie on an
edge of one of the three orthogonal lattice planes through the origin. 
\end{defn} 

As in the pure gauge theory case, the non-trivial global topology of lattices with
periodic boundary conditions leads one to moves that create and destroy structure
on a global scale, in this case lines of charge. 

In the half-unit charge case, a second line needs to be added to absorb the
flux introduced by the first; the ``smallest'' possible global move places a second line
of half-unit charge immediately beside the first. 

In the case of integer valued charge, the lowest energy (and hence smallest change in
amplitude) structure satisfying admissibility is formed by wrapping the smallest possible
cylinder supported by the lattice; because there are $2n$ fundamental irreps,
$n$ can be incident on one side normal to the line, wrap into a cylinder, and enter the 
other to satisfy $c$-edge admissibility.

\begin{defn}[Dimer move] An edge is randomly selected and a single unit of occupancy is
added or removed.  If the result is not consistent with the presence of dimers and cycle
edges (i.e. a cycle that ran through the edge is broken by the removal of occupancy), the move is rejected. 
\end{defn} 

Singly and doubly charged dimers cannot be constructed by plaquette moves. Note a
singly charged dimer contributes $\text{Tr}(U_eU^\dagger_e)=n$ with weight $K^2$ while a doubly charged
dimer contributes $\text{Tr}(U_eU^\dagger_e)\text{Tr}(U_eU^\dagger_e)-\text{Tr}(U_eU^\dagger_eU_eU^\dagger_e)=n^2-n$
with weight $K^4$ (unlike general polymers, we combine the two routings into one configuration). Both types of 
dimers evaluate to constants with respect to the gauge variables, and thus don't couple 
to the gauge bosons. 

\begin{defn}[Junction move] A vertex is randomly selected, and if multiply occupied,
the routing of flux is changed.
\end{defn}
A multiply occupied vertex in the case of a 2-component fermion field has two 
flux paths, which can be routed in two different ways. When a multiply occupied 
vertex first appears as a result of a polymer move, one routing is randomly selected (similarly
in the inverse case, with weightings to preserve detailed balance as discussed above). 
Thus, the junction moves are stricly speaking unnecessary for ergodicity, but may
be used to improve performance of Metropolis algorithm.

\subsubsection{The Algorithm}
Combining the polymer and spin foam moves, given, we give a statement for
a Metropolis algorithm ergodic on the joint ensemble.
\begin{algorithm}
(Joint Fermion-Boson Algorithm). \emph{An iteration of the joint algorithm consists of
choosing one of the seven previously defined moves, which can be organized
as as follows:}
\end{algorithm}
\vspace*{11pt}
\begin{center}
\includegraphics[scale=0.82]{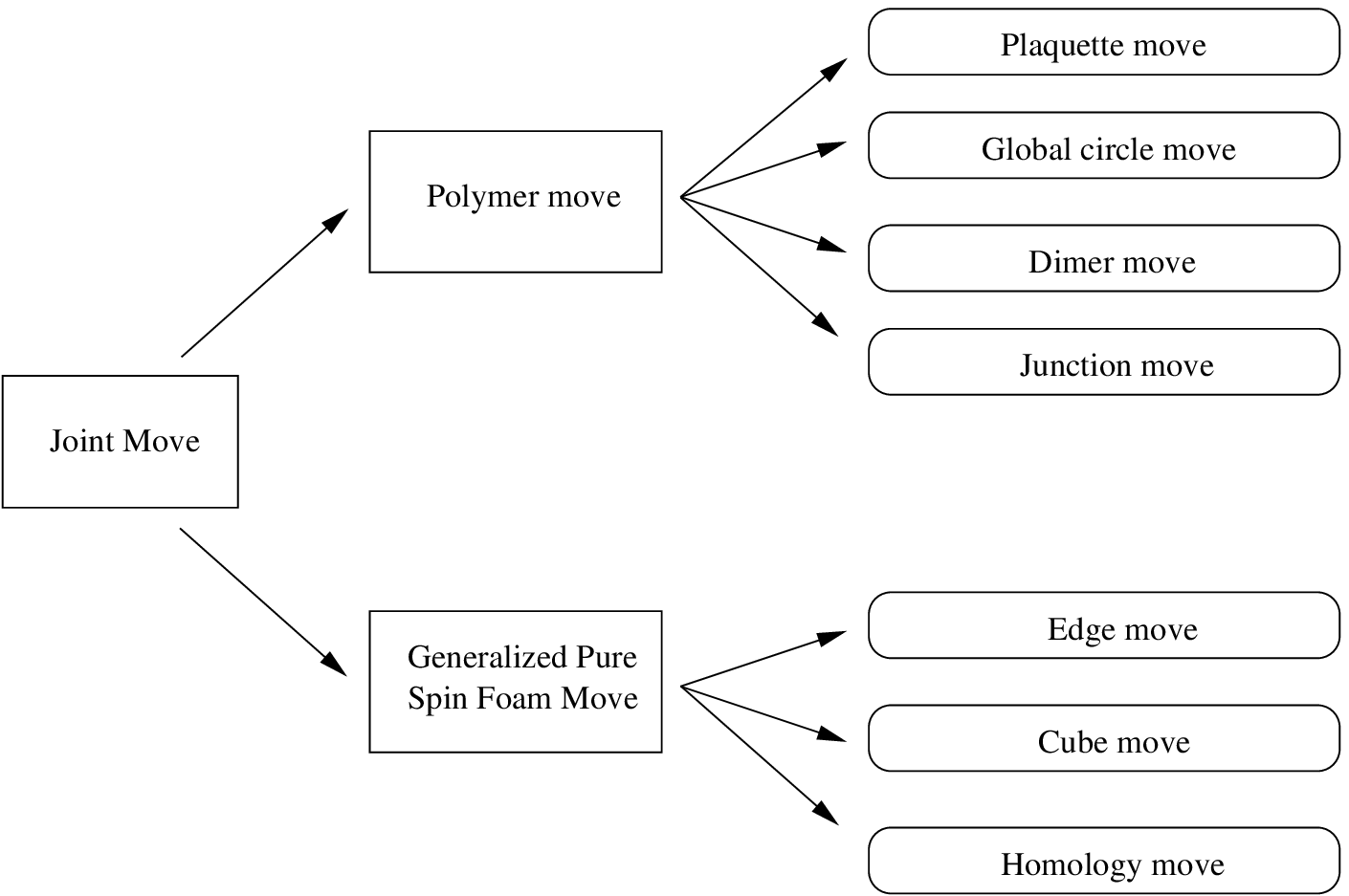}
\end{center}
\vspace*{11pt}
The algorithm can be tuned to improve acceptance rate by adjusting the 
relative frequency of the attempted move types. 

The algorithm also tracks the sign of the configuration changes with 
the creation and destruction of fermion loops, which can occur
with any of the polymer moves. The product of monomer factors $M$ and hopping
factors $K$ are also updated for polymer moves. It is important to emphasize that
changes in both the sign and other factors require only local consideration
of the polymer moves. 

With regard to locality, one sees that the (pure spin foam) homology move
and global circle moves will lead to updates costing on the order of
$L^2$ and $L$ respectively, where $L$ is a characteristic side length
of the lattice. In the pure Yang-Mills case analyzed numerically
in~\cite{CCK}, the homology moves have negligible influence beyond
very small lattice sizes.  It remains to be seen how this is modified
in the joint case, and how large an influence the global circle moves
have. 

Within the scope of the current work, the expectation values of observables depending on dual
degrees of freedom are computed in the usual manner, by averaging the observable 
over the Markov chain generated by the Metropolis algorithm.  
For Wilson loop type observables commonly studied, the expectation value
is actually a ratio of dual charged and dual vacuum partition functions,
with static charge corresponding to the Wilson loop observable present in 
the charged partition function and a vacuum partition function given by
$\mathcal{Z}_J$ of equation~(\ref{eq:jointpartfunc}).  The computation of 
Wilson loop observables of pure Yang-mills and dynamical fermions will be 
reported on in forthcoming work by the author.


\section{Outlook and Conclusions}
We present here a local, exact algorithm for Metropolis simulation
of the fermion-boson vacuum.  The details have been provided for
the case of $D=3$ staggered Kogut-Susskind fermions 
coupled to a Yang-Mills $SU(2)$ field; however the algorithm has a straightforward
generalization to other dimensions and gauge groups.

A limitation of the algorithm as currently given is the species doubling inherent
in the (unrooted) Kogut-Susskind formulation (e.g. in four dimensions there will be four species).
An alternative to Kogut-Susskind fermions which addresses
species doubling was developed by Aroca et al.~\cite{ArocaWorldsheet} and  Fort~\cite{Fort}.
We are currently investigating this modified fermion action in the non-abelian, 
higher dimensional context. Another approach would be to go through a similar procedure 
using Wilson fermions, in which the unwanted doublers become very heavy
in the continuum limit.  

The crucial question for any new fermion algorithm is its performance 
relative to the highly evolved dynamic fermion methods that exist within
the conventional lattice gauge community today.  In three-dimensions, slow-down at weaker
coupling has been observed in recent work on the pure Yang-Mills case~\cite{CCK}.
The situation in $D=4$ is not well understood and is currently the subject of
numerical work by the author, as are improvements in the
original $D=3$ case.   

With regard to the continuum limit, we expect the most critical question for the 
algorithm proposed here is the seriousness of the sign problem. A hard sign problem 
has been discussed as a general feature of the polymer expansion in the continuum 
(small mass) limit~\cite{Montvay90}. While oscillating signs can be overcome
for lattice fermions in certain two dimensional theories~\cite{Montvay90, Wolff}, the author is not aware of 
methods that have successfully addressed the sign problem for $D>2$.
As both the fermion and dual Yang-Mills (spin foam) amplitudes can carry negative signs, an 
important question is how the signs interact; i.e. the problem of signs may be 
harder or easier than for either the free fermion polymer expansion or dual Yang-Mills alone, depending on how
the signs correlate. 

Although numerical developments are required to begin evaluating this proposal, we believe 
the approach may be of considerable interest. We find it 
remarkable that the fermion expansion into polymers and the gauge field dualization 
into spin foams (both of which have been extensively explored on there own), combine 
together in a way that is very compelling geometrically, and allows a local, exact 
Metropolis simulation using gauge-invariant configurations carrying entirely discrete labels.  
    
\begin{acknowledgement*}
The author would like to thank Dan Christensen, Florian Conrady, and Igor Khavkine for 
valuable discussions. The author was supported by NSERC. 
\end{acknowledgement*}

\appendix
\section{Charged $nJ$ Symbols}

\subsection{The dual model with charges}\label{sse:derivation}
In this appendix, we deal specifically with $D=3$, $G=SU(2)$.
Following the discussion in the appendix of~\cite{CCK}, we recall that the dual
partition function (in the absence of charge) has the form
\begin{equation}\label{eq:Zcharexp}
	\mathcal{Z} = \sum_{\{j_p\}} \int \prod_{e\in E} dg_e
		\prod_{p\in P} c_{j_p} \chi_{j_p}(g_p),
\end{equation}
where summation over $j_p$ is over unitary irreducible representations of $SU(2)$.
At this point it is convenient to specialize to a $D=3$ cubic lattice
with periodic boundary conditions; orientation choice is as given
in the appendix of~\cite{CCK}.
With this choice of orientation, the holonomy around a plaquette $p$ is
$g_p = g_1 g_2 g_3^{-1} g_4^{-1}$,
where $g_1, g_2, g_3$ and $g_4$ are the group elements associated to the
edges of the plaquette $p$, starting with an appropriate edge and
going cyclically.
Recall that the inverse $g_i^{-1}$ is used if the
orientation of edge $i$ does not agree with that of $p$.
Thus
\begin{equation}\label{eq:chi}
  \chi_{j_p}(g_p) = U_{j_p}(g_1)^b_a \, U_{j_p}(g_2)^c_b \,
                    U_{j_p}(g_3^{-1})^d_c \, U_{j_p}(g_4^{-1})^a_d \, ,
\end{equation}
where $U_j(g)^b_a$ denotes a matrix element with respect to a
basis of the $j$ representation.  If we insert~\eqref{eq:chi} 
into~\eqref{eq:Zcharexp} and collect together factors depending on
the group element $g_e$, we get a product of independent integrals
over the group, each of the form
\begin{equation}\label{eq:HIdef}
  \int dg_e \, U_{j_1}(g_e)^{b_1}_{a_1} \, U_{j_2}(g_e)^{b_2}_{a_2} \,
               U_{j_3}(g_e^{-1})^{b_3}_{a_3} \, U_{j_4}(g_e^{-1})^{b_4}_{a_4}
  = \int dg_e~\raisebox{-.70cm}{\includegraphics[height=1.8cm]{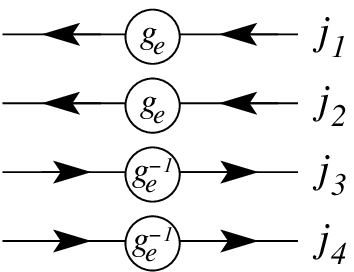}} \,\, .
\end{equation}
Here and below we use a graphical notation for tensor 
contractions, as in~\cite{CCK}.

Equation~\eqref{eq:HIdef} defines a projection operator on the
space of linear maps $j_{4} \otimes j_{3} \ra j_{1} \otimes j_{2}$, so it can 
be resolved into a sum over a basis of intertwiners $I_i: j_{4} \otimes j_{3} \ra j_{1} \otimes j_{2}$
\begin{equation}\label{eq:HIresolution}
\int dg_e~\raisebox{-.75cm}{\includegraphics[height=1.8cm]{cables}} 
	~=~\sum_i \frac{I_i I^*_i}{\langle I^*_i,I_i\rangle}
	~=~\sum_i
		\frac{
			\includegraphics[height=1.5cm]{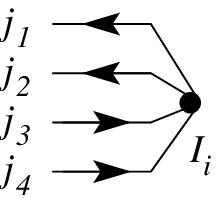} ~~
			\includegraphics[height=1.5cm]{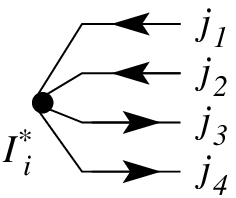}
		} {\raisebox{0cm}{\includegraphics[height=1.7cm]{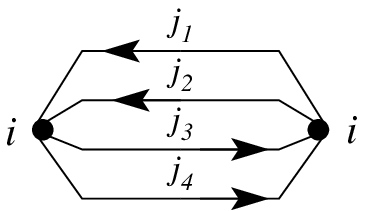}}},
\end{equation}
where the intertwiners 
$I^*_i:j_{1} \otimes j_{2} \ra j_{4} \otimes j_{3}$
are chosen such that the trace $\langle I^*_{i'},I_{i}\rangle$ of the 
composite $I^*_{i'} I_i$ is zero whenever $i'\ne i$ and non-zero if $i'=i$.
The projection property is readily verified.

We next define  $\mathcal{Z}_{\gamma}$, the partition function charged according
to the polymer $\gamma$, as follows

\begin{equation}\label{eq:Zcharexp2}
	\mathcal{Z_{\gamma}} = \sum_{\{j_p\}} \int \prod_{e\in E} dg_e
	\text{Tr}\left( \prod_{e \in \gamma} U^{e}_{c}(g_e)^{i^{+}_e}_{i^{-}_e}\right)
    \prod_{p\in P} c_{j_p} \chi_{j_p}(g_p).
\end{equation}

Collecting matrix factors by dependence on edge variable $g_e$, we find in addition to
the matrices from the four incident plaquettes, a matrix from the edge
$e$ with charge $c$ belonging to the polymer $\gamma$:\footnote{In the case
where an edge is doubly occupied, the integral involves a sixth matrix 
(and the resolving intertwiners an additional input and output arrow). 
It is straightforward to generalize the present analysis to this case; the resulting 
doubly charged $18j$ symbols are shown in Appendix B.} 
\begin{equation}\label{eq:chargededge}
 \int dg_e \, U_{j_1}(g_e)^{b_1}_{a_1} \, U_{j_2}(g_e)^{b_2}_{a_2} \,
               U_{j_3}(g_e^{-1})^{b_3}_{a_3} \, U_{j_4}(g_e^{-1})^{b_4}_{a_4}
               U^{e}_{c}(g_e)^{i^{+}_e}_{i^{-}_e}
=\int dg_e~\raisebox{-.75cm}{\includegraphics[height=1.8cm]{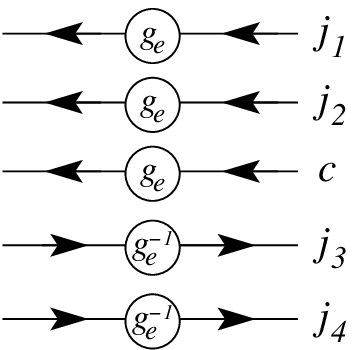}}. 
\end{equation}
As in the pure case, the group integral can be resolved into invariant intertwiners
\begin{equation}\label{eq:HIresolution_c}
\int dg_e~\raisebox{-.75cm}{\includegraphics[height=1.8cm]{cables_charged}} 
	~=~\sum_i \frac{I_i I^*_i}{\langle I^*_i,I_i\rangle}
	~=~\sum_i
		\frac{
			\includegraphics[height=1.5cm]{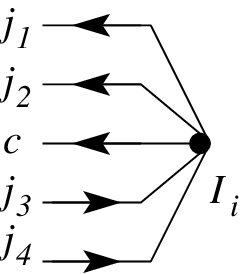} ~~
			\includegraphics[height=1.5cm]{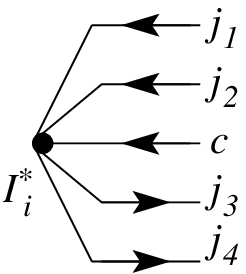}
		} {\raisebox{0cm}{\includegraphics[height=1.7cm]{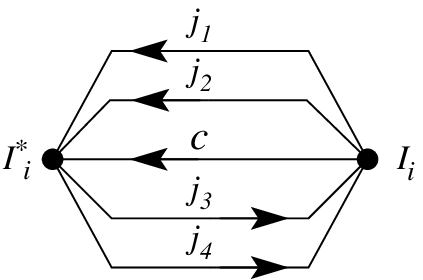}}}.
\end{equation}

If, for each edge of the lattice, we fix a term $i$ in the above
summation, the intertwiners $I_i$ and $I^*_{i}$ can be contracted with those
coming from the other edges, leading to a sum over intertwiner labellings
at every edge. Observe that at edges occupied by polymers, there is more 
than a single intertwiner spin label due to the additional splittings (see 
Figure~\ref{fig:splitv}) introduced by the charge lines. At each vertex of 
the lattice, there will be six intertwiners $I_i$ (some carrying
multiple labels), and their contraction can be graphically represented as an 
octahedral network plus additional lines depending on how the polymer
passes through the vertex. As well, at each edge there will be a normalization
factor corresponding to the denominator of equation~(\ref{eq:HIresolution_c}).

We consider first the vacuum case. In this case, each edge carries only a single 
intertwiner label. The result is the $18j$ symbol central to pure Yang-Mills spin foams, 
\begin{equation}\label{eq:oct}
\begin{matrix}\includegraphics[scale=0.7]{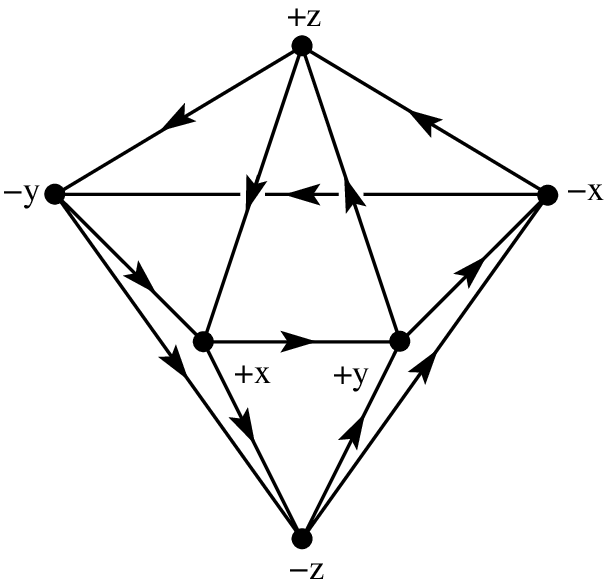}\end{matrix}.
\end{equation}
The vertices are labelled by the directions of the associated
lattice edges emanating from the given lattice vertex, namely 
$\pm x$, $\pm y$, and $\pm z$. 
The value of the $18j$ symbol depends on the choice of basis elements
$I_i$ and $I^{*}_{i'}$ in~\eqref{eq:HIresolution_c}, 
the six summation indices $i$ labelling the edges,
and the 12 incident plaquette labels $j$. 

\begin{figure}[h]
\includegraphics[scale=0.7]{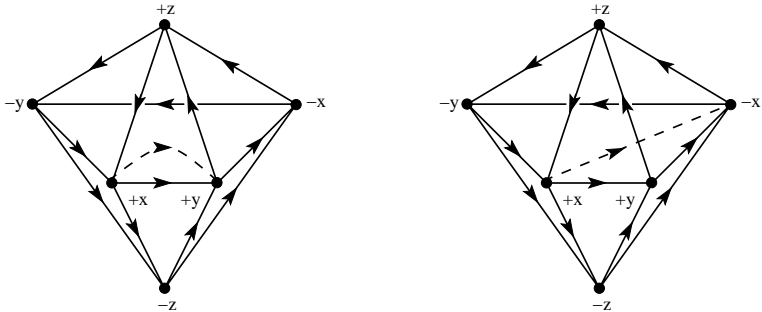}
\caption{Charged $18j$ symbols with flux lines passing through at right
angles (left) and straight through (right).}
\label{fig:oct2}
\end{figure}

\begin{figure}[h]
\includegraphics[scale=0.7]{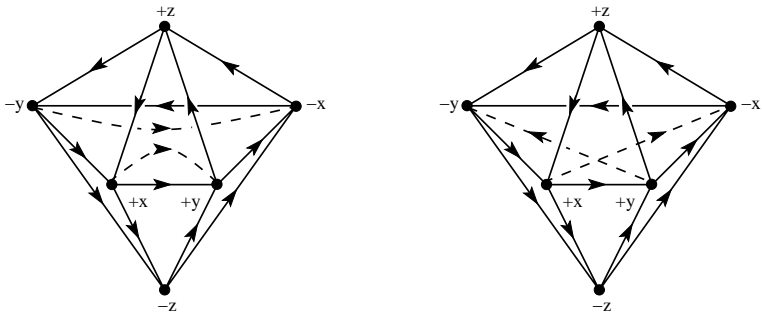}
\caption{Charged $18j$ symbols with two pairs of flux.  Cases with
flux not in the same plane are also possible.}
\label{fig:oct4way}
\end{figure}

We now turn to the case where there is a (single) polymer along one or more
of the edges incident to a vertex. Each charged normalization
 factor $\overline{N} \equiv \langle I^*_i,I_i\rangle$
in the denominator of~\eqref{eq:HIresolution} depends on the charge
$c$ of the fermion at that edge, the intertwiner labels $i$ on that edge, and the labels
of the four plaquettes incident on that edge.
In the presence of external charges, the vacuum $18j$ is modified depending
on whether the line of charge proceeds directly through the vertex or
turns, leaving in a direction perpendicular to entry direction. We call these
cases \emph{charged} $18j$ symbols and denote them by an overline, 
$\overline{18j}(j_v,i_e,\gamma)$ where the additional dependence on gamma
reflects how the polymer charge is routed through the octahedral network. Typical
charged $18j$ symbols are shown in Figure~\ref{fig:oct2}.
Cases where the direction of the arrow on the charged lines is flipped will also 
occur, depending on the polymer orientation. Additionally, as discussed in 
the next section, more than a single line of flux can
pass through a vertex, leading to charged $18j$ symbols of the form shown in 
Figures~\ref{fig:oct4way} and~\ref{fig:18j_2charged}.

In implementing numerical code for this algorithm, a choice of splitting
(grouping of the four plaquettes into $(j_1,j_2)$ and $(j_3,j_4)$ pairs
on opposite sides of the splitting)
is made and each vertex resolved into a 3-valent sub-network with 
up to three non-trivial intertwiner labels, as shown in Figure~\ref{fig:splitv}.
At this point, recoupling moves (see A.2 of~\cite{CCK}, and references therein)
can be used to reduce the spin network to sums and products of know spin networks
such as the $6j$ and theta networks, for which efficient algorithms are available.
It should be noted, however, that different splittings 
lead to differing efficiency in implementation, so some care and experimentation should
be applied to finding an efficient splitting.  Specific splitting schemes
and their performance evaluation will be reported on in forthcoming
numerical work by the author and collaborators.

\section{Polymers with multiply occupied vertices}
In order to couple to the spin foam representation of the gauge theory,
we seek to collect the permutation contributions to the fermion determinant
into traces of products of $U_e$ matrices around closed, oriented loops of
edges.  The case of permutations where a single component is shifted was 
discussed above in Section~\ref{sse:coupled}. For polymers where more than one 
component is shifted at a vertex, recovering a trace formula is somewhat more subtle. 

\begin{figure}
\includegraphics[scale=0.44]{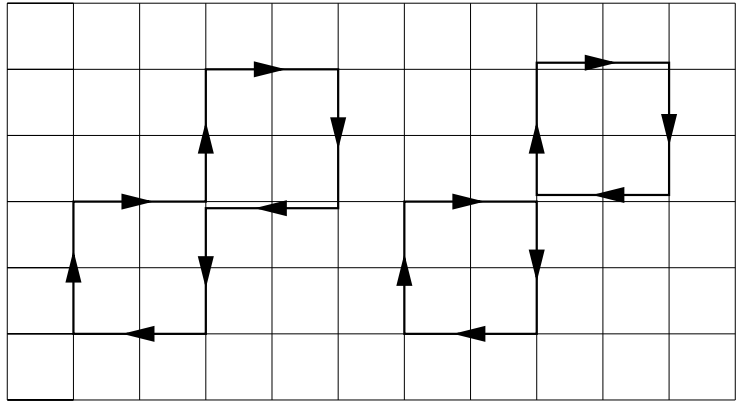}
\caption{Two routings associated with a polymer that self intersects
once at a point.}
\label{fig:pointsplit}
\end{figure}

In Figure ~\ref{fig:pointsplit}, we illustrate a case where a vertex is 
multiply occupied; the orientation is such that there are two possible 
routings that resolve the ambiguity at that vertex. Neither diagram 
by itself corresponds to the desired sum of permutation contributions.
In the matrix multiplications and traces (viewed as a sum over all paths
around a loop), there are terms in each corresponding to paths that are not 
permutations. However, the same undesired terms occur 
with opposite sign in the two diagrams (as one involves paths that form a single
loop, the other paths that lie in two disjoint loops) so the sum of both captures the sum of 
permutations associated with the polymer. 

\begin{figure}
\includegraphics[scale=0.44]{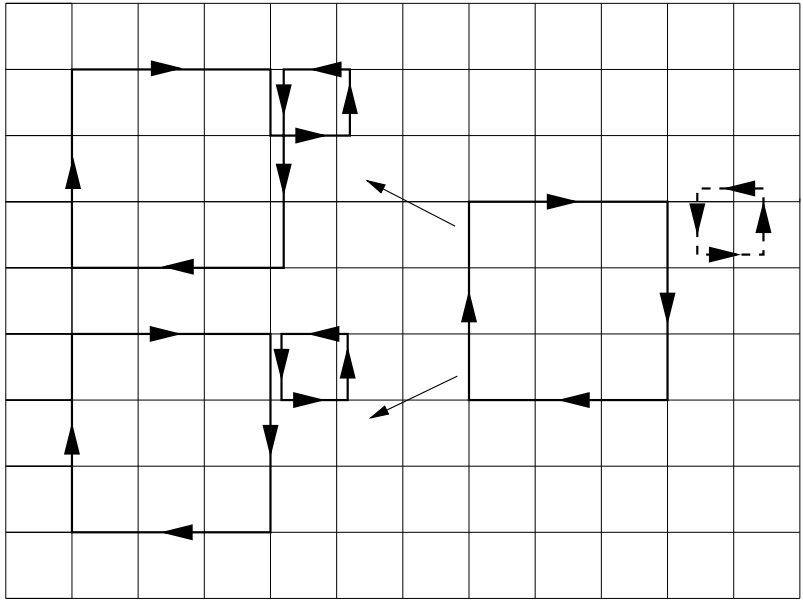}
\caption{A move introducing a doubly occupied edge, for which there
are two distinct routings. }
\label{fig:pchoices}
\end{figure}

\begin{figure}
\includegraphics[scale=0.7]{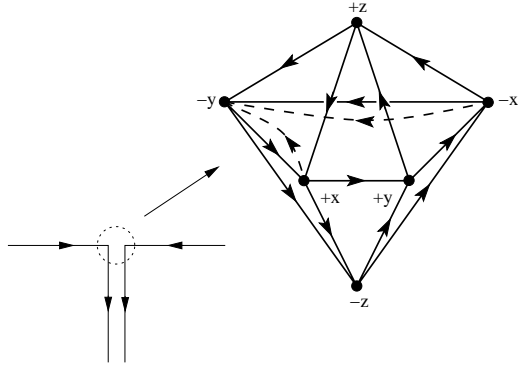}
\caption{A charged 18j symbol containing a 6-valent
node incoming from a doubly charged edge. }
\label{fig:18j_2charged}
\end{figure}

A similar cancellation occurs when two loops share a single edge, i.e.
the edge is multiply occupied. Because there are two multiply ocuppied vertices, 
there are $2^2=4$ routings possible, however only two are topologically 
distinct; two representatives appear in Figure~\ref{fig:pchoices}.  
The cancellation of unphysical paths between the traces over differently 
routed polymers is well known from the hopping parameter expansion (HPE) of
the fermion determinant as discussed for example in~\cite{Rothe}.
As shown in Figure~\ref{fig:18j_2charged}, the presence of a doubly charged
edge leads to a charged $18j$ spin network with a 6-valent node. As well, the
charged normalization factor $\overline{N}$ on a doubly charged edge is as
given in the denominator of equation~(\ref{eq:HIresolution}), but with 
an additional $c$-charged line parallel to the original $c$-charged line.

\end{document}